\def\TR{}
\ifdefined\TR
  \documentclass[12pt,a4paper]{article}
  \usepackage[margin=3.5cm]{geometry}
\else
  \documentclass[article]{jss}
\fi

\ifdefined\TR
  \newcommand{\pkg}[1]{\textbf{#1}}
  \newcommand{\proglang}[1]{\textsf{#1}}
  \newcommand{\code}[1]{\texttt{#1}}
  \usepackage{fancyvrb}
  \DefineVerbatimEnvironment{CodeInput}{Verbatim}{fontshape=sl}
  \DefineVerbatimEnvironment{CodeOutput}{Verbatim}{}
  \newenvironment{CodeChunk}{}{}
  \newcommand{\Prob}{\mathsf{P}}
  \newcommand{\E}{\mathsf{E}}
  \newcommand{\Abstract}[1]{\gdef\savedabstract{#1}}
  \newcommand{\Keywords}[1]{\gdef\savedkeywords{#1}}
  \newcommand{\Plaintitle}[1]{}
  \newcommand{\Shorttitle}[1]{}
  \newcommand{\Plainauthor}[1]{}
  \newcommand{\Plainkeywords}[1]{}
  \newcommand{\Address}[1]{}
  \newcommand{\email}[1]{\texttt{#1}}
  \usepackage{natbib}
\fi

\usepackage{orcidlink,thumbpdf,lmodern}
\usepackage{amsmath,amssymb}
\usepackage[utf8]{inputenc}
\def\..{\,\mathpunct{\ldotp\ldotp}} 

\DeclareUnicodeCharacter{00D7}{\ensuremath{\times}}

\newcommand{\F}[1]{\mathbf{F}_{#1}}
\renewcommand{\Prob}{\operatorname{Pr}}
\DeclareMathOperator{\rank}{rank}

\author{Sebastiano Vigna~\orcidlink{0000-0002-3257-651X}\\Universit\`a degli Studi di Milano\\Dipartimento di Informatica}
\Plainauthor{Sebastiano Vigna}

\title{Modular Rank and Linear--Complexity Tests for Pseudorandom Number Generators}
\Plaintitle{Modular Rank and Linear--Complexity Tests for Pseudorandom Number Generators}
\Shorttitle{Modular Rank and Linear--Complexity Tests}

\Abstract{
  Standard batteries of tests for pseudorandom number generators (such as
  \pkg{dieharder}, the NIST suite, and \pkg{TestU01}) provide two empirical
  tests for linearity, the binary rank and linear-complexity tests. Both operate
  over the field $\F{2}$, and thus detect generators that are linear over
  $\F{2}$. However, generators can be linear over a larger field, as in the case
  of congruential generators, single-modulus multiple-recursive recurrences, and
  of matrix generators such as MIXMAX. We introduce a \emph{modular} version of
  the rank and linear-complexity tests, and provide \pkg{modlin}, a
  \proglang{Rust} program that implements it efficiently for fields of prime
  size. \pkg{modlin} can detect in minutes statistical bias in all current
  CERN's \pkg{ROOT}'s implementations of the MIXMAX generator, for which no
  standard statistical test failure has been reported before.
}

\Plainkeywords{matrix rank, linear complexity, finite fields, pseudorandom number
  generators, empirical testing, MIXMAX, Rust}

\Address{
  Sebastiano Vigna\\
  Dipartimento di Informatica\\
  Universit\`a degli Studi di Milano\\
  Via Celoria 18\\
  20133 Milano, Italy\\
  E-mail: \email{sebastiano.vigna@unimi.it}\\
  URL: \url{https://vigna.di.unimi.it/}
}

\ifdefined\TR
  \usepackage{hyperref}
\fi

\begin{document}

\ifdefined\TR
  \maketitle
  \begin{abstract}
  \savedabstract

  \end{abstract}
\fi


\section{Introduction} \label{sec:intro}

Empirical testing subjects the output of a pseudorandom number generator (PRNG)
to a battery of statistical tests, each designed to expose a particular kind of
regularity. Examples include the \pkg{dieharder} battery \citep{dieharder}, the
NIST suite \citep{Rukhin+others:2010}, and \pkg{TestU01}
\citep{LEcuyer+Simard:2007}.

Two tests, in particular, look for \emph{linear structure}: the \emph{binary rank
	test}, introduced by \citet{MarTMSRNS}, and the (binary) \emph{linear-complexity
	test}~\citep{CarALC,ErdETBK,Gustavson:1976}. The binary rank test fills a matrix with bits produced by the generator
and computes its rank over the binary field $\F{2}$; the linear-complexity test
measures using the Berlekamp--Massey algorithm \citep{Massey:1969} the length of
the shortest linear recurrence the output of the generator obeys. The two are intimately
related, as a generator whose bit stream satisfies a short linear recurrence
over $\F{2}$ fails both: it produces rank-deficient matrices, and a complexity
far below the random value. Together they are the standard way to find bias in
linear-feedback shift registers and in $\F{2}$-linear generators (e.g.,
\code{xorshift}~\citep{MarXR}, the Mersenne Twister~\citep{MaNMT}, etc.), all of which are linear over
$\F{2}$.

A limitation of these tests is that they
work only over $\F{2}$. They are therefore unable to find bias in a generator that carries no $\F{2}$
structure but is linear over a different field. Rank tests over larger
fields have been considered, but always in characteristic two:
\citet{Grosek+Vojvoda+Krchnavy:2009} build a $\chi^2$ rank test over $\F{q}$ for
$q = 2$ and $q = 4$ and use it to expose linear-feedback shift registers that the
binary tests accept. But such work remains tied to the structure of
characteristic-two arithmetic.

Generators exhibiting a linear structure over large fields are not exotic. A
single multiple-recursive recurrence modulo a prime $p$ is linear over $\F{p}$;
and matrix generators \citep{Niederreiter:1992} evolve a vector state by a
linear map over $\F{p}$. The popular combined generators that splice together
several such recurrences over distinct primes, such as
\code{MRG32k3a}~\citep{l1999good}, lie outside this class. The linear
structure of multiple-recursive and matrix generators is classically probed by
the \emph{spectral} and \emph{lattice} tests
\citep{Coveyou+MacPherson:1967,Knuth:1998,LEcuyer+Couture:1997}: figures of
merit, computed from the recurrence itself, that gauge the regularity of the
lattice formed by successive outputs; low figures of merit can sometimes lead
to failure in collision and birthay-spacing tests. The tests we develop instead work
empirically, on the output stream alone.

A prominent recent example of the single-field case is given by MIXMAX
\citep{Savvidy:2015}, adopted in the CERN's \pkg{ROOT} framework: it iterates an
$N \times N$ integer matrix over the Mersenne prime field $\F{2^{61}-1}$, with
$N$ as large as $240$ or $256$. Its bit stream exhibits no low-order $\F{2}$
recurrence, so the binary rank and linear-complexity tests pass it; yet over
$\F{2^{61}-1}$ it is, by construction, exactly linear. The \pkg{ROOT}
documentation claims that MIXMAX has ``proof random properties''
\citep{ROOT:TRandom}, and that it passes the BigCrush test from
TestU01~\citep{ROOT:MixMaxEngine}. \citet{LEcuyer+Wambergue+Bourceret:2019}
have analyzed MIXMAX using spectral techniques and have been able to find some
structure detectable by collision tests on very specific subsets of coordinates,
but the current versions of MIXMAX in CERN's \pkg{ROOT} have been patched to
avoid these issues (e.g., by not using the first value of each iteration, and by
decimating the output) and do not currently fail any standard collision or
birthday-spacings test.

In this paper, we discuss the extension of the rank and linear-complexity tests
from the binary field to any finite field, and describe \pkg{modlin}, a tool
written in \proglang{Rust} \citep{Matsakis+Klock:2014} that implements such tests
efficiently (the implementation is for the time being limited to
fields $\F{p}$ with $p$ prime and $p < 2^{63}$). Our main motivation is to show
that, contrary to the case of collisions and birthday spacings, MIXMAX fails both the rank and the
linear-complexity tests over $\F{2^{61}-1}$.

The only input either test needs, beyond the output of the generator itself, is
the field modulus: a sufficiently large matrix will be rank deficient, and a
sufficiently long stream will have too low complexity. Indeed,
for a generator that is linear over $\F{p}$ the recovered value is exactly the
output linear complexity.

While computing the linear complexity is relatively cheap ($O(n^2)$), the rank
test is more expensive ($O(n^3)$ field multiplications). Our implementation uses
a blocked Gaussian elimination, parallelized across cores, and uses
division-free modular reduction, making it possible to run rank tests on
matrices of dimension well above $100\,000$, if sufficient memory is available, in a
few hours. The corresponding linear-complexity test runs in minutes on a single
core for a sequence of the same length.

Section~\ref{sec:tests} defines the two statistics and explains why a modular
linear generator is detected by both. Section~\ref{sec:null} develops their null
distributions and the $p$-values. Section~\ref{sec:impl} describes the
implementation, Section~\ref{sec:usage} shows how to use \pkg{modlin}, and
Section~\ref{sec:results} reports results on the MIXMAX generators, showing how
the statistical defects we detect can influence actual computations; we give concluding remarks
in Section~\ref{sec:conclusions}.

\pkg{modlin} is available under the Apache License~2.0 or the GNU Lesser
General Public License~2.1 or
later.\footnote{\url{https://github.com/vigna/modlin-rs/}}


\section{The two tests} \label{sec:tests}

Fix a prime $p$ and let $x_0, x_1, x_2, \dots$ be the output sequence of the
generator under test, each $x_i$ reduced to an element of the field
$\F{p}$ (Section~\ref{sec:impl} describes the mapping used and its effect on
uniformity).

For the \emph{modular rank test} one chooses a matrix dimension $n$ and forms the $n
	\times n$ matrix \begin{equation} \label{eq:matrix} M = \bigl(\, x_{\,i\,n + j}
	\,\bigr)_{0 \le i,j < n}, \end{equation} that is, row $i$ is the block of $n$
successive outputs $x_{in}, \dots, x_{in+n-1}$. The test statistic is the rank
of $M$ over $\F{p}$.

For the \emph{modular linear-complexity test} of length $n$ one takes an output sequence
$x_0, \dots, x_{n-1}$ and computes its \emph{linear complexity}
$L_n$ over $\F{p}$, that is, the length of the shortest linear-feedback shift register
that generates it, or, equivalently, the order of the shortest linear recurrence
that generates it, using the Berlekamp--Massey algorithm \citep{Massey:1969}. The test
statistic is $L_n$.

Now, let us say that a generator is \emph{linear over $\F{p}$ with state dimension $k$} if it
maintains a state $s \in \F{p}^{k}$ that evolves by a linear map,
$s \mapsto A s$ with $A \in \F{p}^{k \times k}$, and emits, at each step, one or
more values that are fixed linear functions of the state.

Suppose first that the generator emits a \emph{single} value per step, so that
$x_i =  v\bigl(A^{i} s_0\bigr)$ for a fixed linear map $v: \F{p}^k\to \F{p}$ and initial state
$s_0$. Such a sequence satisfies a linear recurrence of order at most $k$ over
$\F{p}$ (i.e., its \emph{linear complexity} is at most $k$) because $A$ satisfies
its own characteristic polynomial, of degree $k$. Consequently every length-$n$
window of the sequence is determined by its first $k$ entries through the
recurrence, so all windows lie in a subspace of $\F{p}^{n}$ of dimension at most
$k$. The rows of $M$ in \eqref{eq:matrix} are such windows, hence
\begin{equation} \label{eq:rankbound}
	\rank M \le k,
\end{equation}
and as soon as $n > k$ the matrix is \emph{rank deficient}: $\rank M \le k < n$.

Linear complexity works analogously: since the stream obeys a recurrence of
order at most $k$, we have $L_n \le k$, far below the value $\lceil n/2 \rceil$
a random stream exhibits (Section~\ref{sec:null}) for sufficiently large $n$.

What if the generator emits more than one value per state step, as a vector or
matrix generator does? In that case, reading $b$ values per step interleaves $b$
sequences, each of linear complexity at most $k$, so the output stream is itself
a recurrence over $\F{p}$ of linear complexity $L \le bk$. Both bounds then hold
with $L$ in place of $k$.

For example, in the case of MIXMAX-$N$, which emits $N-1$ values per step from
an $N$-dimensional state, the output stream has linear complexity at most $L = N(N-1)$.

Note that the two tests are sensitive to the same quantity but with different
tradeoffs. Computing the linear complexity is cheap, but the result is
\emph{fragile}: it requires the \emph{entire} stream to obey one recurrence, so
a single value off the recurrence forces the complexity to jump back toward
$n/2$. A generator that is only partly linear (e.g., linear just in some
outputs, or interleaved with other values) may well pass a linear-complexity
test.

On the contrary, usually the rank test is more \emph{robust}: for a sufficiently
large matrix, a rank-deficient core might keep the rank below full even if the
entries are scrambled or interleaved with other values (e.g., a
\code{xoroshiro128+}~\citep{BlVSLPNG} generator will fail binary rank tests for
sufficiently large matrices, even if its output is scrambled nonlinearly by a
sum, as its low bits are of low linear complexity).


\section{The null distributions and the \texorpdfstring{$p$}{p}-values} \label{sec:null}

Under the hypothesis that the generator emits independent values uniform over
$\F{p}$, both statistics have classical exact distributions. We state these
results for a general field of size $q$ (any prime power); in
our setting $q = p$.

\subsection{The rank distribution} \label{sec:nullrank}

The $n^2$ entries of $M$ are independent and uniform, so $M$ is a uniform random
$n \times n$ matrix over $\F{q}$. Its rank distribution is known
\citep{Landsberg1893}: the probability that an $n \times n$
matrix over $\F{q}$ has rank $\rho$ is
\begin{equation} \label{eq:rankdist}
	\Prob[\rank = \rho] = q^{-n^2}\,
	\prod_{i=0}^{\rho-1} \frac{(q^{n}-q^{i})^2}{q^{\rho}-q^{i}},
\end{equation}
and in particular the matrix has full rank with probability
\begin{equation} \label{eq:fullrank}
	\Prob[\rank = n] = \prod_{i=1}^{n} \bigl(1 - q^{-i}\bigr).
\end{equation}
The deficiency probability $\Prob[\rank < n] = 1 - \prod_{i=1}^{n}(1-q^{-i})$ is,
for large $q$, dominated by its first term: a square matrix over $\F{q}$ is
singular with probability
\begin{equation} \label{eq:defprob}
	\Prob[\rank < n] = q^{-1} + q^{-2} - q^{-5} - \cdots \approx q^{-1},
\end{equation}
and more generally the \emph{corank} $n - \rank M$ is at least $d$ with probability $\approx q^{-d^2}$.
For a large prime such full rank is essentially certain, and any rank
deficiency is strong evidence against the randomness hypothesis.

\subsection{The linear-complexity distribution} \label{sec:nulllc}

The linear complexity $L_n$ of a uniform random length-$n$ sequence over $\F{q}$
has a classical exact distribution~\citep{Gustavson:1976}:
$\Prob[L_n = 0] = q^{-n}$, and for $1 \le L \le n$,
\begin{equation} \label{eq:lcpmf}
	\Prob[L_n = L] = (q-1)
	\begin{cases}
		q^{\,2L-n-1}, & 1 \le L \le \lfloor n/2 \rfloor, \\
		q^{\,n-2L},   & \lfloor n/2 \rfloor < L \le n,
	\end{cases}
\end{equation}
a distribution sharply peaked at the mode $\lceil n/2 \rceil$, where
$\Prob[L_n = \lceil n/2 \rceil] = 1 - 1/q$. Its mean is $n/2$ up to an $O(1/q)$
correction~\citep{Gustavson:1976}:\footnote{Note that in~\citet{Gustavson:1976} $E(n)$ equals $2\,\E[L_n]$.}
\begin{equation} \label{eq:lcmean}
	\E[L_n] = \tfrac{n}{2} + \frac{q}{(q+1)^2}
	+ \frac{(n \bmod 2)\,(q-1)^2}{2(q+1)^2} - O(n\,q^{-n}),
\end{equation}
which tends to the mode $\lceil n/2 \rceil$ as $q \to \infty$.

Over a
large prime field the complexity of a random sequence is $\lceil n/2 \rceil$ with
probability $1 - 1/q$, and falling by a factor $q^{-2}$ for every step away from
the mode. So, exactly as with the rank, any complexity lower than the mode
is strong evidence against the randomness hypothesis.


\section{Implementation details} \label{sec:impl}

\pkg{modlin} is written in \proglang{Rust}~\citep{Matsakis+Klock:2014}. The
generator is selected at compile time, so the generation loop is fully
specialized and inlined. Included with the tool are \proglang{Rust} ports of the
three versions of MIXMAX from \pkg{ROOT}~\citep{Savvidy:2015} and a reference
\code{xoroshiro128++} generator~\citep{BlVSLPNG}.

Field arithmetic is performed for an arbitrary prime $p < 2^{63}$, so that a
single addition cannot overflow a $64$-bit word; products are taken in $128$-bit
arithmetic and reduced. Modular reduction uses precomputed $128$-bit
reciprocals. It would be easy to extend the implementation to primes up to
$2^{127}$, but the cost of the field operations would increase significantly.
The same considerations hold for extending the implementation to powers of primes.

Each generator output is mapped to a field element by reduction modulo $p$. In
the case of MIXMAX, this reduction is exact (i.e., a bijection). When instead
the generator emits $w$-bit words and $p$ does not divide $2^{w}$, as for the
$64$-bit control \code{xoroshiro128++}, plain reduction is not perfectly
uniform: the $2^{w}\bmod p$ smallest residues are over-represented by a relative
factor $1/\lfloor 2^{w}/p\rfloor$, which is small when $p$ lies well below
$2^{w}$. Crucially, and independently of $p$, this perturbs only the marginal
distribution and introduces no $\F{p}$-linear relation among the outputs, so it
can never by itself cause rank deficiency or lower the linear complexity. Exact
uniformity can be obtained for any $p$ by rejecting the $2^{w}\bmod p$ largest
words; for the prime used in our experiments, $p = 2^{61}-1$, the bias is in any
case only $\approx 2^{-61}$ in total variation.

The rank of an $n \times n$ matrix is computed by parallel blocked Gaussian
elimination to row-echelon form, using $n^2$ $64$-bit memory locations and
$O(n^3)$ field operations \citep{vonzurGathen+Gerhard:2013}. The linear
complexity is computed by the Berlekamp--Massey algorithm~\citep{Massey:1969} in
$O(n^2)$ field operations and $O(n)$ space.

For each sample the tool reports a $p$-value: $\Prob[\text{corank} \ge c]$
for a matrix of corank $c$, and $\Prob[L_n \le \ell]$ for a sequence of
complexity $\ell$. For a non-anomalous sample this value is close to $1$:
exactly $1$ for a full-rank matrix, and at least $q/(q+1)$ (tending to $1$ as
$\ell$ grows toward $n$) for a complexity $\ell$ at or above the mode. For an
anomalous sample it is $\approx q^{-c^2}$ (respectively, $\approx
	q^{\,2\ell-n}$), so the tests are one-sided. Note that if $p$ is sufficiently large,
the $p$-value underflows to zero for almost all anomalous samples, and in that
case the tool reports as the $p$-value $10^{-307}$, the smallest representable power of $10$.

Typically, in the binary version of the test one considers multiple matrices and
computes a Pearson $\chi^2$ test over the resulting histograms, because over a
small field the rank or complexity distribution spreads across many
well-populated cells. This is not very useful in the case of large fields, but
\pkg{modlin} makes it possible, in any case, to repeat the test a given number
of times: the user can then combine the resulting $p$-values as they see fit
(e.g., using a coarse Bonferroni correction).


\section[Using modlin]{Using \pkg{modlin}} \label{sec:usage}

\pkg{modlin} is a command-line program written in \proglang{Rust} and
distributed in source form through its GitHub repository: because testing a new
generator means adding it to the code (see the end of this section), there is no
precompiled binary or published crate. The generator under test is selected at
compile time through a Cargo feature (i.e., conditional compilation), so each
build tests one generator; a single invocation runs one test. The program
provides features \code{xoroshiro128pp}, \code{mixmax17}, \code{mixmax} (a
synonym of \code{mixmax240}), and \code{mixmax256}, plus the output modifier
\code{mixmax-lux}, which applies RANLUX-style decimation (keep~$5$, discard~$9$)
to a MIXMAX generator and is combined with one of the \code{mixmax} features.

A run is launched with \code{cargo run -r -F <generator>} (here \code{-r}
abbreviates \code{-{}-release} and \code{-F} abbreviates \code{-{}-features}),
followed by \code{-{}-} and the test options. Exactly one of \code{-R}~\textit{side}
(modular rank test on \textit{side}\,$\times$\,\textit{side} matrices) and
\code{-L}~\textit{length} (modular linear-complexity test on sequences of the
given length) must be given. The modulus is set with \code{-p}~\textit{prime}
(any prime below~$2^{63}$); an optional positional argument gives the number of
repetitions (default~$1$), \code{-S}~\textit{seed} sets the generator seed
(default~$0$), and \code{-{}-log-interval} tunes progress logging. The number of
threads for the parallel rank computation is controlled by the environment
variable \code{RAYON\_NUM\_THREADS}, and logging verbosity by \code{RUST\_LOG}.

As an example, both tests detect the bias of the $17$-dimensional MIXMAX
generator of \pkg{ROOT} over $\F{2^{61}-1}$ in a few milliseconds. The
linear-complexity test recovers the exact complexity $N(N-1) = 272$:
\begin{CodeChunk}
\begin{CodeInput}
cargo run -r -F mixmax17 -- -L 1000 -p 2305843009213693951 -S 1
\end{CodeInput}
\begin{CodeOutput}
Generator: MIXMAX (TRandomMixMax17, N=17)
Seed: 0x0000000000000001
Running a modular linear-complexity test using 0.031 MiB of RAM: [...]
[...]
Sequence 1/1	linear complexity=272	p=1e-307
\end{CodeOutput}
\end{CodeChunk}
while the rank test, on a $1000\times1000$ matrix, finds a corank of $966$:
\begin{CodeChunk}
\begin{CodeInput}
cargo run -r -F mixmax17 -- -R 1000 -p 2305843009213693951 -S 1
\end{CodeInput}
\begin{CodeOutput}
Generator: MIXMAX (TRandomMixMax17, N=17)
Seed: 0x0000000000000001
Running a modular rank test using 7.629 MiB of RAM: [...]
[...]
Matrix 1/1	corank=966	p=1e-307
\end{CodeOutput}
\end{CodeChunk}
In both cases the per-sample $p$-value underflows and is reported as the smallest
representable power of ten, $10^{-307}$. By contrast, the control
\code{xoroshiro128++} produces a full-rank matrix (corank~$0$, with $p$-value $1$) under the
same parameters, and likewise passes the linear-complexity test; so does every
MIXMAX generator when tested over a prime other than $2^{61}-1$.

To test a generator not
already included, one adds a feature in \code{Cargo.toml} and a corresponding
implementation, exposing a \code{next\_u64} step function, in the \code{prng}
module.


\section{Results} \label{sec:results}

We applied both tests to the three MIXMAX generators of CERN's \pkg{ROOT}, all
linear over $\F{2^{61}-1}$, with state dimensions $N = 240$
(\code{TRandomMixMax}), $17$ (\code{TRandomMixMax17}), and $256$
(\code{TRandomMixMax256}); the last generator additionally discards two of every three
iterations to improve quality. Each emits $N-1$ values per step (the first value of the tuple
was found to be particularly problematic by \citet{LEcuyer+Wambergue+Bourceret:2019}),
so their output stream has linear complexity $N(N-1)$ (see
Section~\ref{sec:tests}).

As control we use \code{xoroshiro128++}~\citep{BlVSLPNG}, a scrambled linear
generator. Finally, we also test \code{TRandomMixMax17-Lux}, a variant of
\code{TRandomMixMax17} based on decimation \emph{\`a la}
RANLUX~\citep{LUSCHER1994100} (we keep $5$ values and discard $9$) that
highlights the robustness tradeoff of the two tests.

\begin{table}[t]
	\centering
	\begin{tabular}{lrr rr rr}
		\hline
		                           &     &           & \multicolumn{2}{c}{rank test} & \multicolumn{2}{c}{complexity test}                          \\
		generator                  & $N$ & $N(N-1)$  & $n$                           & corank                              & $n$        & $L_n$     \\
		\hline
		\code{TRandomMixMax17}     & 17  & 272       & $1000$                        & $966$                               & $1000$     & $272$     \\
		\code{TRandomMixMax}       & 240 & $57\,360$ & $80\,000$                     & $22\,640$                           & $200\,000$ & $57\,360$ \\
		\code{TRandomMixMax256}    & 256 & $65\,280$ & $80\,000$                     & $66\,944$                           & $200\,000$ & $65\,280$ \\
		\code{TRandomMixMax17-Lux} & 17  & 272       & $1000$                        & $983$                               & $1000$     & $500$     \\
		\code{xoroshiro128++}      & --- & ---       & $1000$                        & $0$                                 & $1000$     & $500$     \\
		\hline
	\end{tabular}
	\caption{Results of the modular rank and linear-complexity tests for $p=2^{61} -1$ on the three MIXMAX generators of CERN's \pkg{ROOT}, a decimated variant (\code{TRandomMixMax17-Lux}), and the control \code{xoroshiro128++}.}
	\label{tab:results}
\end{table}

Table~\ref{tab:results} collects both tests. The generated random matrices are
defective, and the modular linear-complexity test recovers the exact complexity
$N(N-1)$ of every MIXMAX generator, provided the sequence is long enough: the
Berlekamp--Massey algorithm pins down a complexity $L$ only once it has seen
about $2L$ values, so the linear-complexity test uses lengths above $2N(N-1)$,
whereas the rank test already exposes the deficiency at the smaller side
$n = 80\,000$. The $p$-value computation underflows, so \pkg{modlin} reports
a $p$-value of $10^{-307}$.

With the same parameters, \code{TRandomMixMax17-Lux} can fool the
linear-complexity test, but not the rank test. The control generator
\code{xoroshiro128++} passes both tests. As a further check, we ran the test for
some other values of $p$, for which all generators pass both tests.

The linear-complexity tests and the small rank tests take less than a minute. The two
large rank tests require 48\,GiB of RAM and a few hours of computation.

\subsection{An instructive example} \label{sec:example}

The bias we detect is not a mere artifact of our tests: like the
$\F{2}$-linearity of the Mersenne Twister, the $\F{2^{61}-1}$-linearity of
MIXMAX leaves a visible trace in ordinary computations that involve no field
arithmetic at all. We give here an example in the spirit of the one discussed
by \citet{VigHTLGMT} for the binary case.

Consider a researcher who simply wants to study random integer matrices---say,
some statistical property of their characteristic polynomials---using MIXMAX to
fill their entries. She computes a batch of characteristic polynomials (over the
integers, rationals, or reals---that is immaterial) and proceeds to use them as
if they were a sample of random integer polynomials from matrices with integer
coefficients chosen at random from a range.

\begin{figure}[t]
\centering
\includegraphics{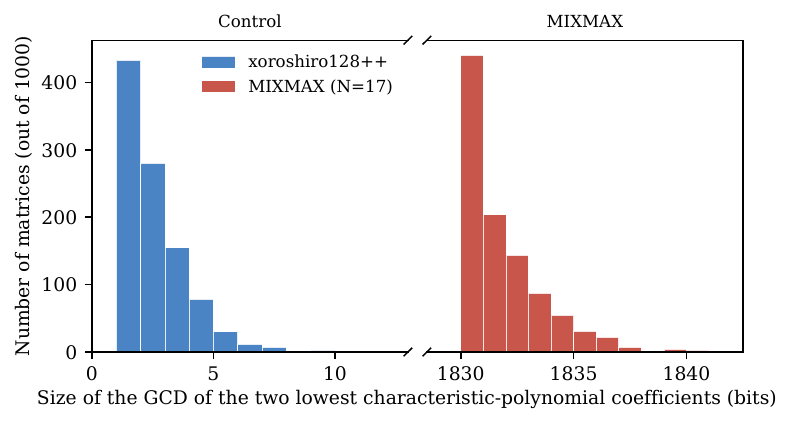}
\caption{\label{fig:charpolygcd}The distribution of the size (in bits) of the
greatest common divisor of the two lowest-order coefficients of the integer
characteristic polynomial of a thousand $48\times48$ matrices filled with the
output of \code{TRandomMixMax17} and of the control \code{xoroshiro128++}. Note the
broken horizontal axis: the two histograms are separated by some eighteen
hundred bits.}
\end{figure}

But are they? If they were random, the greatest common divisor of their
lower-order coefficients (i.e., the determinant and the sums of principal
minors) would be small, those coefficients behaving like unrelated large
integers. We put this to the test with \code{TRandomMixMax17} and, as a control,
\code{xoroshiro128++}, whose $64$-bit output we reduce modulo $2^{61}-1$ and
shift by one, so that both generators emit integers in the same range
$[1 \.. 2^{61})$. We fill a thousand $48\times48$ matrices
from each, compute the integer characteristic polynomial of each matrix, and
record the number of bits of the greatest common divisor of its two lowest-order
coefficients; Figure~\ref{fig:charpolygcd} shows the two distributions.

For \code{xoroshiro128++} the greatest common divisor is a handful of bits: the
coefficients are essentially coprime, as one would expect from random integers.
For MIXMAX, instead, the two lowest-order coefficients share a common factor of
about eighteen hundred bits.

Our researcher would be witnessing the $\F{2^{61}-1}$-linearity of MIXMAX
influencing her results even though no computation over $\F{2^{61}-1}$ is
involved. The mechanism is the one already at work in Table~\ref{tab:results}:
her matrices have low rank over $\F{2^{61}-1}$, so their characteristic
polynomial over the field is divisible by a large power of $x$ and lacks its
low-order coefficients; and the coefficients over the field are simply the
reduction modulo $2^{61}-1$ of the coefficients over $\mathbf{Z}$. The
coefficients that vanish modulo the prime are therefore exactly the integer
coefficients that are divisible by (powers of) it, which explains the
preposterous size of the GCD (about $30\log_2 p$ bits, where the multiplier is
roughly the corank of the matrices).

\section{Conclusions} \label{sec:conclusions}

We have presented \pkg{modlin}, a \proglang{Rust} statistical tool for testing
pseudorandom number generators. \pkg{modlin} implements new modular versions of the
binary rank and linear-complexity tests, making it possible for the first
time to detect bias in generators that are linear over a large prime field, such
as the MIXMAX family.

The tests detect linearity over the specific prime field supplied; a generator
linear over a different modulus, or genuinely nonlinear, will pass. In particular,
\emph{combined} generators assembled from components over distinct prime moduli,
such as \code{MRG32k3a}~\citep{l1999good}, will pass the test for any of the
component moduli.

In spite of this specificity, however, \pkg{modlin} provides clear statistical
evidence that the MIXMAX generators, which are proposed by CERN in the
\pkg{ROOT} framework as ``proof random'', are biased; in fact, as our example
shows, this bias can creep into computations in quite insidious ways. This is
even more relevant in view of the fact that the authors of MIXMAX have already
applied a number of patches to the original MIXMAX proposal, such as skipping
the first value of each iteration, or decimating the output, to solve issues
previously reported by~\citet{LEcuyer+Wambergue+Bourceret:2019}, and yet, these
patches cannot hide the linear statistical bias of the generator.

Considering the importance of unbiased generation in large-scale simulations, we
can only hope that CERN switches to stream ciphers such as AES
in counter mode~\citep{SMDPRN}, which, as an additional bonus, on modern
hardware are significantly faster than the flawed MIXMAX.

\ifdefined\TR\else
\section*{Replication materials}

The experiments in this article are fully reproducible. A single driver
\code{replicate-all.sh} reproduces every empirical result: it runs
\code{replicate-tests.sh}, which regenerates every entry of Table~\ref{tab:results} and
the field-specificity checks, and it builds the self-contained directory
\code{charpoly-gcd}, which regenerates Figure~\ref{fig:charpolygcd} using
\proglang{Sage} for the exact characteristic-polynomial arithmetic. All runs use
a fixed seed, and a transcript of a complete run is provided in
\code{replication-output.txt}. These materials accompany this article together
with the \pkg{modlin} source.
\fi

\bibliography{modlin}

\end{document}